%% file: Moth_astroph.tex
\documentclass[preprint]{aastex}
\slugcomment{To appear in Astron. J.}
\shortauthors{Moth \& Elston}
\shorttitle{Color Gradients and Surface Brightness Profiles}

\begin{document}

\title{Color Gradients and Surface Brightness Profiles of Galaxies in the Hubble
Deep Field-North}

\author{Pimol Moth \altaffilmark{1} \& Richard J. Elston
\altaffilmark{2}}
\affil{Dept. of Astronomy, University of Florida, 211 Bryant Space 
Science Center, P.O. Box 112055, Gainesville, FL  32611}

\altaffiltext{1}{NASA Florida Space Grant Fellow}

\altaffiltext{2}{Presidential Early Career Award for Scientists and Engineers Recipient}

\begin{abstract}
We fit elliptical isophotes  to the Hubble Deep Field-North  WFPC-2 and NICMOS
data to study the rest-frame $(UV_{218}-U_{300})_o$
color
profiles and rest-frame B surface brightness profiles of 33 intermediate redshift galaxies ($0.5 \leq z \leq 1.2$)
with $I_{814}$ $<$ 25 and 50 high redshift galaxies 
($2.0 \leq z \leq 3.5$) with $H_{160}$ $<$ 27. 
From the weighted least-squares fit to the color profiles we find that, at
intermediate redshifts, the galaxies possess 
negative color
gradients ($\langle$ $\Delta(UV_{218}-U_{300})_o$/$\Delta$log(r) $\rangle $= $-0.091$ 
 $\pm$ 0.007 mag dex$^{-1}$) 
indicating a reddening towards the center of the profile similar to
local samples whereas, at
high redshifts, the galaxies possess positive color gradients ($\langle$ $\Delta(UV_{218}-U_{300})_o$/$\Delta$log(r)
$\rangle $= 0.272 $\pm$ 0.007 mag dex$^{-1}$) indicating
that star formation is
more centrally concentrated.  Although the presence of dust can cause some reddening to occur towards the
centers of the profiles seen at intermediate redshifts, it can not explain the strong central blueing of light   
seen at high redshifts.  
Thus, we are witnessing a population of galaxies with strong positive color
gradients at high redshifts which do not seem to exist 
in large numbers at
lower redshifts.  This indicates that star formation is more centrally concentrated in the distant galaxy sample
which differs from the prevalent mode of extended disk star formation that we observe in the local universe.     
Additionally, we find that it is critical to correct for PSF effects when evaluating the surface brightness 
profiles since at small scale lengths and faint magnitudes, an $r^{1/4}$ profile can be smoothed out substantially to
become consistent with an exponential profile.  
After correcting for PSF effects, we find that at higher look-back time, the fraction of
galaxies possessing
exponential profiles have slightly decreased while the fraction of galaxies possessing $r^{1/4}$ profiles have slightly increased.
Our results also suggest a statistically insignificant increase in the fraction of peculiar/irregular type galaxies. 
We compare our results with recent semi-analytical models which treat galaxy formation and evolution following the
cold dark matter hierarchical framework.
\end{abstract}


\section{Introduction}

Two of the most fundamental and intriguing questions in astronomy are 
how galaxies form and how they evolve with time.  In order to 
answer these questions, we must be able to compare and contrast the 
properties of galaxies at different redshifts.  
Whereas we have a wealth of information about galaxies at z $\leq$ 1.0, our
knowledge of the properties of galaxies at higher redshifts is limited.
However, with the advent of the Hubble Deep Field-North (HDF-N) 
project \citep{wil96,thom98,dic99,dic00}, we can 
probe to fainter surface
brightness limits and smaller angular scales than before, the sizes,
shapes, and colors of distant galaxies which will ultimately yield
important clues to understanding their structure, formation, and
subsequent evolution.  The near-infrared 
data from NICMOS combined with the optical data from WFPC-2  
give us the unique opportunity to compare the surface brightness properties 
of galaxies at the same rest-frame wavelengths
over a range of redshifts 
allowing us to better understand galaxy formation and evolution from an
observational standpoint.

One observational test of galaxy evolution is the study of color profiles which will give us an idea of the distribution
of stellar populations in the
galaxies and whether this distribution changes with time.
Due to limits in resolution, 
the study of the color profiles of galaxies have been predominantly restricted 
to the low and intermediate redshift regimes.
However, the deep, high resolution, multi-wavelength data from the
HDF-N probed in this study provides us with the rare opportunity to study
the color profiles 
of galaxies at higher redshifts than those studied in the past.  
Previous studies of early-type galaxies at z $\leq$ 1.0 have shown that
they tend to have redder colors in
their central regions and gradually become bluer outwards \citep{tamu00,tam00,vad88,fran89,pel90}.  
This trend in the color profiles can be
explained by either a stellar age or metallicity gradient which become degenerate at
$z = 0$ \citep{silva94}.
Most studies concur that 
models involving the metallicity gradient best reproduce the color 
gradients seen in early-type galaxies at z $\leq$ 1.0.
On the other hand, the gradients observed 
in late-type galaxies may be due to
both age and metallicity effects, i.e., the central parts
of the galaxies in general have older stars and higher metallicity than the
outer parts making them have redder colors towards the centers.
In their study of the near-IR and optical color profiles of 86 face-on disk
dominated galaxies, \citet{deJ96} concluded that their color gradients were
best reproduced by models involving both stellar age and metallicity gradients.
The existence or lack of dust in galaxies can also complicate matters and
must be addressed when interpreting color profiles since, in theory, dust may also be responsible for the central
reddening in
galaxies if we assume that dust generally tends to be more concentrated in
the center and consequently would produce more extinction there \citep{deJ96,evans94,buy94}.  

A study of the color profiles and surface brightness profiles of galaxies spanning a wide range of
redshifts
will help us place constraints on galaxy formation and evolution by enabling us
to compare what we learn from observations with what we predict from
theoretical models.  
Currently, the hierarchical 
structure formation model \citep{bau98,cole94,kauf97,rouk97,whit91} represents the popular 
framework for how structure formed and evolved in the universe.  This 
model assumes 
that the  universe is dominated by nonbaryonic dark matter which only 
interacts with visible matter through its gravitational influence and 
ultimately determines where galaxies will form.  It 
predicts that the gravitational perturbations in the early universe will 
cause the smallest mass fluctuations to collapse first and then to 
subsequently merge into progressively larger structures until they form 
the mature galaxies we observe today. \citet{bau98} (hereafter BCFL) analyzed the properties of the high redshift Lyman break 
galaxies in the context of this model.  In their semi-analytic treatment of 
galaxy formation in hierarchical clustering theories, they generated mock 
catalogs of the high redshift Lyman break galaxies (LBGs) using the color criterion 
imposed by \citet{steid93} and modeled the growth of dark matter haloes 
by the accretion of matter through mergers taking into account the 
cooling of gas into stars and feedback.  \citet{SPF00} (hereafter SPF) also applied semi-analytic
models of 
galaxy formation within the hierarchical clustering framework to analyze the population of Lyman break galaxies at high redshift.
Since they conveniently address
the properties of
Lyman break galaxies (the sample of significant interest to us) within
the current popular framework for galaxy
formation and evolution, we will interpret our results within the context
of the BCFL and SPF models.  

\section{Data and Sample Selection}
It is important to study the properties of the galaxies at 
different redshifts at the same rest-frame wavelengths because some galaxies 
might experience what is called "morphological k-corrections".  
This is a phenomena which 
occurs when galaxies at different redshifts look different when viewed 
through the same passband since the passband would represent
different rest-frame wavelengths. For example, the $I_{814}$ image for a 
galaxy at redshift $z = 0.5$ would reflect the properties of the galaxy at 
the rest optical wavelengths whose light would include both young and old  
stellar populations, whereas for a galaxy at redshift $z = 3$, the $I_{814}$ 
image would represent the rest-UV wavelengths of the 
galaxy which is dominated by the younger, bluer stellar populations.
Thus, if we want to compare the properties of the same stellar 
populations, we 
need to choose our passbands such that they reflect the same rest 
wavelengths at all redshifts.  

The original HDF-N data was a Director's 
Discretionary program on HST designed to image a field at high galactic 
latitude using the WFPC-2 camera in four passbands spanning the 
wavelengths $0.3-0.8$ $\mu$m \citep{wil96}.  
At low redshifts, these images were sufficient to study the galaxies in 
the rest-frame optical wavelengths, however at z $>$ 1, the WFPC-2 images
could only be used to study the rest ultraviolet properties of the 
galaxies since their rest optical wavelengths were shifted into 
the near-IR.  Near-IR data of galaxies in the HDF-N were 
available from ground-based observations \citep{hog97}.  
However, since 
the ground-based data were taken at much lower resolution ($\sim$ 1 $\arcsec$) and reached 
shallower depths than did the WFPC-2 images, it was very difficult to 
compare the  properties of high redshift galaxies with their lower redshift 
counterparts at the same rest-frame wavelengths.        
To address this problem, \citet{dic00}, embarked on a General 
Observer's (GO) Program 7817 to image the 
entire WFPC-2 region of the HDF-N in the near-IR using the NICMOS Camera 3 F110W 
and F160W filters.  The data reached an average depth of $AB = 26.1$ in F110W and F160W 
with a signal-to-noise ratio of 10 in 0.$\arcsec$7 diameter aperture and has a PSF of $FWHM = 0.$\arcsec$22$ 
after the dithered data was combined by drizzling. 
To avoid the morphological 
k-correction problem, we study the intermediate redshift galaxies in the HDF-N using 
the WFPC-2 
images taken in F450W ($B_{450}$), F606W ($V_{606}$) and F814W ($I_{814}$)
and the high redshift galaxies using WFPC-2  F814W ($I_{814}$) and 
the NICMOS CAM3 images taken in F110W ($J_{110}$) and F160W ($H_{160}$) in 
order to probe their rest-frame near-UV and optical properties. 

Since we will be comparing the color profiles of galaxies at 
a range of redshifts, it is important to establish what we mean by intermediate
and 
high redshift galaxies.   We will define \it intermediate redshift\rm \ 
galaxies 
as those having redshifts between 0.5 $\leq$ z $\leq$ 1.2  and \it high 
redshift \rm \ galaxies as those having redshifts between 2.0 $\leq$ z 
$\leq$ 3.5. We do not include galaxies 
with 1.2 $\leq$ z $\leq$ 2.0 in our sample because at these redshifts, 
there are no prominent optical features to 
constrain the redshifts with the high precision required in our 
study.  Our sample consists of 33 intermediate redshift galaxies with $I_{814}$ $<$ 25 identified by 
spectroscopic redshifts \citep{coh96,zepf97} and 50 high redshift galaxies with
$H_{160}$ $<$ 27, of which 25 were 
identified by spectroscopic redshifts \citep{steid96,low97}
and 25 by photometric redshifts \citep{bud00}.  In our paper, we assume a cosomology where 
 $Ho = 71 km/s/Mpc$ and $q_o = 0.3$.  

\section{Methodology}

\subsection{Surface Photometry}
We performed detailed surface photometry on our sample of galaxies using 
the ELLIPSE task in the IRAF/STSDAS package.  
This task fits elliptical isophotes to 
the galaxies using the iterative scheme described in \citet{jed87}.  
We entered inital guesses for the ellipse center (\it x,y\rm), ellipticity 
(\it e\rm), and 
position angle ($\phi$) and allowed the task to update these parameters 
following 
the iteration scheme.   The errors in the azimuthally 
averaged intensity are obtained 
from the rms scatter of the intensity measurements along each fitted 
isophote.  In order to generate color profiles, it was important that we 
kept the ellipses fixed between the different filters since positional
displacement of the isophotes in the different filters may cause the 
color distribution of  the galaxy to be artificially asymmetric.  
Thus, we combined 
the images taken in different passbands, fitted ellipses to the 
combined image, 
and then used the set of ellipses generated from the combined image as 
the input ellipses for no-fit mode on each individual image. In this way, we
made sure that the surface brightness profiles from the different bands were
generated using the same fitted ellipses.  It is also important that the PSFs
between the different filters match.  This was accomplished by degrading the  
$B_{450}$, $V_{606}$, $I_{814}$, and $J_{110}$ images to
match the resolution of the $H_{160}$ image.  The PSF matching was verified by looking for color gradients in stellar
profiles. Many of the 
galaxies, especially at high redshifts, possessed nearby companions.  In 
order to perform the fits only on the galaxy of interest, we masked out the 
companion.  We then converted all fluxes to AB magnitudes.

\subsection{Fitting Surface Brightness Profiles}
From the elliptical isophotes, we produced azimuthally averaged radial 
surface brightness profiles.  For our intermediate redshift sample, we produced surface 
brightness profiles of the images taken in  
$I_{814}$, while for the high redshift sample, we produced surface 
brightness profiles for the images taken in
$H_{160}$.  At the respective redshifts of the galaxies, these images 
represent approximately their rest-frame B wavelengths.  

We analyzed the resulting intensity distribution using the STSDAS 
task NFIT1D to perform $\chi^2$ fitting to the surface 
brightness profiles.  Normally we would fit the profiles with the sum of an $r^{1/4}$ bulge \citep{deV48}
 and an exponential disk \citep{fre70} profile.  However, for most of the 
galaxies, especially at high redshifts, the scale lengths of the 
galaxies were so small and severely affected by PSF smearing that it was 
difficult to obtain reliable fits using these profiles. Thus, we decided to  
fit the radial light profiles with a generalized exponential \citep{ser68}:
\begin{equation}
S(r)=S(e)exp[-(1.9992n-0.3271)(r/R(e)^{1/n}-1)]
\end{equation}
where \it R(e)\rm \ is the half-light radius, \it S(e)\rm \ is the flux at \it 
R(e)\rm, $S(r)$ is the surface brightness at r, and 
\it n\rm \ is the S\'ersic index which tells us about the degree of 
flattening in the profile ($n = 1$ recovers 
the exponential disk profile and $n = 4$ recovers the $r^{1/4}$ law).    
\citet{truj01} have applied an analytic approach to
study the effects of the PSF on the S\'ersic profiles.  They found that, of the free parameters in the S\'ersic
profile, the one that is affected most by the PSF is $n$ which is smaller in the observed profile than in the
actual profile.   Furthermore, the higher the original value of $n$ is, the more it will be affected by
the PSF. 
At the faint magnitudes and small scale lengths of typical high redshift galaxies in the HDF-N, we
will show that
the PSF can potentially 
flatten both an exponential and an $r^{1/4}$ profile into a S\'ersic 
profile with \it n\rm \ $<$ 1.  
It then becomes challenging to distinguish
between 
galaxies having exponential profiles from those having de Vaucouleurs 
profiles.  The next section explains our attempts to understand the 
effects of the PSF on the surface brightness profiles in order to more accurately 
classify morphologies.

\subsection{Creating Model Galaxies to Account for PSF Effects} 
In order to account for the effects of the PSF on the data, we created
model galaxies and 
convolved them with the PSF to match the resolution of the data.  
To reproduce the surface brightness profiles of the intermediate redshift sample,  
we created 10 galaxies with exponential disk profiles ($\it n\rm = 1$) and 
10 with $r^{1/4}$ ($\it n\rm = 4$) bulge profiles with varying  half-light radii
and $I_{814}$ magnitudes = 21, 22, 23, and 24 (this spans the range of magnitudes in our
intermediate redshift sample) for a total of 80 galaxies. Since the WFPC-2 and NICMOS images have all been
degraded to match the resolution
of the $H_{160}$ passband, we convolved 
the model galaxies 
with an isolated, unsaturated star taken from this 
passband.  
We performed the convolution using the IRAF task 
FCONVOLVE which takes 
the fourier transform of the model galaxy and the PSF (normalized to 
conserve flux), multiplies them together, and then takes the inverse 
transform of the product to produce the final convolved galaxy.        
We then added the model galaxies onto blank areas of the sky in the HDF $I_{814}$ image. 
By doing so, we include the contribution of the background noise which exists in the image.  However, we have chosen to
neglect the
poisson
noise from the galaxies themselves since it will be negligible compared to the background noise.  
To simulate our high redshift sample, we
repeated the process but added the model galaxies onto blank areas of the HDF $H_{160}$ image with 
$H_{160}$ magnitudes = 23 and 24.  

We then generated surface brightness profiles and fit S\'ersic profiles to the convolved disk and 
spheroid galaxies in order to determine the degree of flattening due to the 
PSF.  Tables 1 and 2 lists the half-light radii and the S\'ersic indices of the convolved model galaxies 
with a range of magnitudes in the
$I_{814}$ and $H_{160}$ 
passbands respectively.  From Tables 1 and 2, 
we can see that after convolution with the 
PSF, the profiles of all the galaxies are significantly flattened, 
i.e. their S\'ersic index $n$ decreased significantly from the original profile 
(recall that $n = 1$ for disks and $n = 4$ for spheroids).  In fact, Table 1 shows that 
after convolution with the PSF, the S\'ersic indices of the mock intermediate redshift galaxies originally possessing an $r^{1/4}$ profile can
drop all the way to a value of 1 which is indicative of an exponential profile.  Table 2 shows an even more drastic
drop in S\'ersic indices of the mock high redshift galaxies after they had been convolved with the PSF of the HDF.  At these small
scale lengths and
faint magnitudes typical of the high redshift galaxies in our sample, the S\'ersic indices of the model spheroids drop to
1 and below rendering them almost indistinguishable from the model disks.
Furthermore, the
smaller the 
galaxy is, the more the value of $n$ decreases for both the disk and 
bulge models.  Consequently it would be very 
difficult to determine from the convolved profile whether it originally possessed
an exponential or an $r^{1/4}$ profile.  

Fortunately, trends 
exist which allow us to reliably distinguish between the two types of galaxies even after 
they have been significantly flattened by the PSF.   
We notice that at a given radius and magnitude both profiles are severely smoothed out, but that $n$ for the spheroid is
always 
higher than $n$ for the disk, i.e. the disk has a more flattened profile.   
For example, at $I_{814}$ = 21 and half-light 
radius of approximately 0.$\arcsec$84, $n$ for the spheroid is 1.736
whereas $n$ for the disk is 0.789 (Table 1) which is significantly lower.
This trend occurs at all radii and magnitudes.  
Thus, although the surface brightness profiles of the galaxies in the HDF-N are 
significantly affected by the PSF, we can still broadly classify galaxies 
morphologically by comparing the value of $n$ and the half-light radii  
of these galaxies at a given magnitude with those of our model galaxies. 
However, caution must be taken when using
this method to
classify the morphologies of galaxies with half-light radii less than $0.12\arcsec$ since it would  
be pushing the resolution limit of the instrument and, consequently, the results would not be
reliable.
Fortunately, there are no galaxies in our sample where this is the case.  We must also note that in our high redshift
sample, we can only reliably classify the morphologies of galaxies with magnitudes less than $H_{160}$ = 24.5
since, at fainter magnitudes, the S\'ersic indices of a disk and spheroid at a given radius is within the range of
the scatter.  We must keep in mind also that our galaxy classification scheme is not intended to be used to classify
galaxies on a case by case basis, but rather to
obtain statistics about the galaxy population as a whole.   
  	
\subsection{Generating Color Profiles and Determining the Gradients} 
We determined the color profiles of the galaxies by subtracting the 
surface brightness profile of one passband from the surface brightness profile of
the adjacent passband.  For our 
intermediate redshift sample, we generated $(B_{450}-V_{606})$
color profiles, and for our high redshift samples, we produced
$(I_{814}-J_{110})$ color profiles. We then applied k-corrections described in the following
section to transform them to their
rest-frame $(UV_{218}-U_{300})_o$  colors where UV $\sim$ 2192 \AA \ and U $\sim$ 2943 \AA.  
The k-corrections applied are minimal since we have chosen to work with passbands which
closely map to their rest-frame wavelengths. 
We then performed a weighted least-squares fit to the profile.  We started the fit 0.14\arcsec 
\ out from 
the center of the profile since the color differences near the center 
might reflect slight differences in the PSF between the two passbands 
rather than indicate true color changes in the galaxy. From the fits, we 
were able to obtain the value of the color gradient 
($\Delta(UV_{218}-U_{300})_o$/$\Delta$log(r)) 
where $(UV_{218}-U_{300})_o$ is the rest-frame color and r is the radius in 
kpc. The main source of error in this 
procedure is the uncertainty in the determination of the local sky background values.
The  global sky background subtraction might be adequate for the 
larger and brighter galaxies in our intermediate redshift sample, but for the  
smaller and
fainter galaxies in our high redshift sample, there could be residuals in 
the background which could significantly change the value and even the sign of 
the color gradient if the effect is huge.   We 
determined the value of the local sky background by extending the elliptical isophotes out 
past five scale lengths in 0.12\arcsec \ increments taking care to mask out nearby
objects.  We then adopted the value of our background as the typical 
value of the isophotes located far enough from the object such that the flux
level ceased to decrease steadily, but instead fluctuated around some value.  From this method, 
we found that in general there had been an overestimation in the 
global background value in the reduced images used for this study.
                    
\subsection{K-corrections}
When we observe two galaxies at different distances with the same 
intrinsic brightness through the same passband, we will measure two 
different magnitudes not only because of the inverse square law, but also 
because their spectral energy distributions (SED) will be shifted towards 
longer wavelengths and also "stretched" by a factor of $(1+z)$ due to the 
expansion of the universe.  In order to correct for this effect and   
to convert the color profiles to their 
rest-frame values, we need to apply k-corrections to our results. We 
use the definition of the k-correction from \citet{oke68}: 
\begin{equation}
\ K_i(z)=2.5log_{10}(1+z)+2.5log_{10}\bigg[\int 
S_i(\lambda)F(\lambda)d(\lambda)\bigg/\int S_i(\lambda)F(\lambda/1+z)d(\lambda)\bigg]  
\end{equation}
where $S_i(\lambda)$ is the sensitivity function of the detector i, 
$F(\lambda)$ is the observed energy flux 
density, $F(\lambda/1+z)$ is the energy flux density of the galaxy at rest, 
$d(\lambda)$ is the bandwidth of the detector \it i\rm, 
and $K_i(z)$ is the k-correction of the 
galaxy observed through detector \it i\rm \ at redshift z.  The first 
term of the 
correction is due to the "stretching" of the passband whereas the second 
term represents the shifting of the SED to longer 
wavelengths. 

Since k-corrections do not currently exist for the NICMOS passbands, we 
generated 
our own set by using the SEDs from \citet{dev99}. They 
used the STARDUST model to 
produce SEDs for 17 galaxies ranging from local spirals, starbursts, 
luminous infrared galaxies, and ultraluminous infrared galaxies 
(ULIRGS) extending from the far-UV to sub-mm wavelengths.  Their SEDs also 
included the effects of internal extinction and emission of dust in the 
galaxies.  

\section{Surface Brightness Profiles}

In the process of analyzing the color profiles of the galaxies, we also studied their surface brightness
profiles. 
Tables 3 and 4 list the statistics for the 33 galaxies in the intermediate redshift and 50 galaxies in the high redshift
regime respectively.  
Column (4) of Tables 3 and 4 lists the S\'ersic indices in the $I_{814}$ passband for
galaxies at intermediate redshifts and the $H_{160}$ passband for galaxies at high redshifts.  
These
passbands translate to approximately the rest-frame B
band
for all the galaxies. 
Ideally
galaxies which possess exponential profiles would have a S\'ersic index of 1 whereas galaxies which possess a de Vaucouleurs
profile would have a S\'ersic index of 4.  Thus, from a naive inspection of the S\'ersic indices of the galaxies at high
redshifts, we would assume that only one (ID 4-555.11 with $n$ = 1.467) has a steeper profile than an
exponential.  
However, we have demonstrated in section 3.3 that when we classify morphologies based on the S\'ersic indices, it is critical to
take into account the role of the PSF since in many cases, especially at high redshifts, the PSF can smooth what was
originally an $r^{1/4}$ profile to one with $n$ $<$ 1.   The 5th column lists our
profile classification for each galaxy (after taking into account PSF effects) based on the galaxy models discussed in
section
3.3 where "deV" represents a de Vaucouleurs $r^{1/4}$ profile, "Exp" represents an exponential
profile, "Int" represents an intermediate profile with 1 $\leq$ $n$ $\leq$ 4 (these galaxies most likely possess both a bulge and disk 
component), and "O" represents "Others"
for galaxies showing irregular structure which could not be fit with any profile.  
In our attempt to classify the morphologies of the galaxies, we do not automatically assume
that objects which possesss exponential profiles are disks while objects which possess de Vaucouleurs profiles are spheroids
since this is not necessarily always the case because many dwarf spheroids have been shown to possess exponential
profiles \citep{koo94,koo95}.
We naively lump into one category ("O") those galaxies which have irregular profiles and do not follow an exponential or
$r^{1/4}$ law. \citet{wu99} explain that there are actually
two types of "irregular" galaxies. One type is the
traditional irregular which is the late-type system classified in the Hubble scheme whereas the other type possesses an
irregular profile as a result of galaxy-galaxy interactions and mergers.  However, it is not within the scope of this paper
to differentiate between the two types.  Thus, it is evident that our classification scheme represents an oversimplification
designed to obtain a sense of the overall trends in the morphologies of the galaxies and
should not be taken as a robust classification scheme on the individual level. 

Figures $1-3$ show the surface brightness profiles and the fits to the profiles of all the galaxies in our sample.  The
Williams et
al.
(1996) identification numbers
and S\'ersic indices
are included in the plots. Figures 4 and 5 represent plots of the half-light radii versus the S\'ersic indices ($r_{1/2} - n$) of the model disks and
spheroids after they had been convolved with the HDF PSF in the $I_{814}$ (with $I_{814}$ = 21, 22, 23, and 24) and $H_{160}$
(with $H_{160}$ = 23 and 24) passband respectively.   
Again, for the high redshift sample, the results from the galaxy models are
only reliable for $H_{160}$ $\leq$ 24.5 since for fainter magnitudes, it is difficult to clearly distinguish the locus of S\'ersic indices
for the model spheroids from the locus of S\'ersic indices for the model disks.  
The triangles represent the values for the model spheroids and the asterisks represent the values for the model disks.  The
dashed and dotted lines represent weighted least-squares fit to the model spheroids and model disks 
respectively, whereas the solid lines
represent the
1$\sigma$ error in the fits.  For all $I_{814}$
magnitudes and for $H_{160}$ $\leq$ 24.5, we have also
plotted the values of the galaxies in our intermediate and high redshift sample (represented by squares) 
to attempt to classify their morphologies by observing where they fall on the diagram.  
We do not include in these plots those galaxies which could not be fit by the S\'ersic law since their
profiles were too irregular.  These galaxies were, however, still included in our statistics and were labeled as "O" for "Others".     
At intermediate redshifts, the morphologies are better constrained since the galaxies are brighter
and the distinction between the model spheroids and disks is more apparent.  As a further check on the accuracy of our classification
system, we can utilize the images we have of the intermediate redshift galaxies since, for many, they are large enough and bright enough to classify
their morphologies by eye.  

With the $r_{1/2} - n$ plots, we have attempted to classify the morphologies of all the galaxies in the intermediate redshift bin and those with
$H_{160}$ $\leq$ 24.5 in the high redshift bin, although it should be kept in mind that the process is rather crude
since it relies
mainly on the galaxy models which should be regarded as highly speculative at small scale lengths and faint magnitudes.  
The galaxies whose S\'ersic indices fall above the locus for the model spheroids were classified as posessing
$r^{1/4}$ profiles since this would 
indicate that they possess profiles that are steeper than our models, whereas those whose
S\'ersic indices fall below the locus for the model disks are classified as possessing exponential profiles since this
would indicate they possess
profiles that are shallower than our models.  For the galaxies which have S\'ersic indices in between the locus of our model spheroids and
disks, we placed an acceptance limit of 5$\sigma$.  If the values of the S\'ersic indices were less than 5$\sigma$ away from the locus of spheroids, we
classified them as possessing $r^{1/4}$ profiles, if they were less than 5$\sigma$ from the locus of the disks,
we classified them as possessing exponential profiles, and if they were more than 5$\sigma$ away from either loci, we classified
them as possessing intermediate profiles.  There were a few galaxies
which possessed S\'ersic indices with less than 5$\sigma$ deviations from both
loci.  In these cases, we classified them according to the magnitude of their deviation.  If the deviation from the locus of model spheroids was less
than it was from the locus of model disks, then we classified those galaxies as possessing $r^{1/4}$ profiles.  If however, the converse was true,
we classified those galaxies as possessing exponential profiles.   But here is where we introduce some uncertainty in our classification scheme.  Since
these galaxies have such small deviations from both loci,
even though we associate these galaxies with a certain profile, it does not mean that it can not possess the other types of profiles.  For example, ID
3-404 was classified as possessing an exponential profile since its S\'ersic index was only 2.1$\sigma$ away from the locus of the model disks,
however, there is still a good chance that it could possess an $r^{1/4}$ profile since it was only 3.3$\sigma$ away from the locus of model
spheroids.  Or it could also possibly possess an intermediate profile.  There are only two galaxies in the intermediate
redshift sample (IDs 2-353 and 3-404) and one in the high redshift sample (ID 2-321) which have less than 5$\sigma$ deviations from
both loci.  If we include these uncertainties, it will at most increase the poissonian errors by 1 percent.  Thus, the profile
classification is very robust, and the uncertainties in the classification system is dominated by the poissonian error.  
     
Table 5 summarizes our best effort to classify the morphologies of the galaxies based on the combination of the results from
the fits to the surface brightness profiles of the galaxies in the two redshift bins, the plots of the galaxy
models, and visual inspection.  
Quoted are the percentages found for each profile with their poissonian error. The number of
galaxies possessing that profile is in parenthesis. 
In our intermediate
redshift sample, out of 32 galaxies, we classified 28 $\pm$ 9\% as possessing $r^{1/4}$ profiles, 53 $\pm$ 13\% as possessing
exponential profiles,
13 $\pm$ 6\% as
intermediates,
and 6 $\pm$ 3\% as "Others" since they possessed irregular substructure.  The galaxy that we
did not include in the statistics
is ID 3-610 in which both an $r^{1/4}$ and and exponential 
component could be resolved.  The S\'ersic index for the inner region of this galaxy is 2.32 while for the outer, disk
part it is 0.826. This is
the only galaxy large enough in our sample to distinguish the two components.
In our high redshift sample only 20 galaxies had $H_{160}$ $\leq 24.5$.  Out of these galaxies, we
classified 60 $\pm$ 15\%
as possessing $r^{1/4}$ profiles, 
20 $\pm$ 10\% as possessing exponential profiles,  and 20 $\pm$ 10\% as
"Others".  In the high redshift bin there are no galaxies which we classified as intermediates.  This does not mean that no intermediates exist at
these redshifts, but rather that since the separation between the locus of model spheroids and model disks is very small, there is not a
clearly defined region for the intermediate cases.   
   
We must stress that we are limited by small number statistics, especially at high redshifts since we are only able to classify 
the
morphologies of 20 out of our original 50 galaxies.  Nevertheless, we can still comment on the general
trends seen.  From Table 5, we see that, at higher look-back time, the fraction of the galaxies possessing $r^{1/4}$ profiles
increased by 32 $\pm$ 18\% whereas the fraction of exponential-type profiles decreased by 33 $\pm$ 16\%.    
These differences are marginally significant since they represent $\sim$ 2$\sigma$ changes in both cases.
Our results differ from what has been previously published
\citep{mar98}.  By using a
profile decomposition method on 522 galaxies down to $m_{F814W}$ $<$ 26, \citet{mar98} found that
the majority of HDF-N galaxies are disk
dominated with only 8\% having dominant bulge fractions.  They interpreted their results as measuring a decrease in the
number of bulge dominated galaxies as a function of look-back time.  Given the high degree of difficulty in the 
classification of the morphologies of these small faint galaxies, which are severely affected by the PSF and the limitation of
small number statistics, it is not
surprising
that at this point, we are not able to reach a general consensus.  
At high redshifts, we also see a statistically insignificant increase of 14 $\pm$ 10\% in the fraction of galaxies which we have classified as
"Others".  This is along the lines of what was observed in previous studies of galaxies in the HDF-N which 
have shown that by $I_{AB}$ $>$
24, there are
few galaxies possessing traditional Hubble sequence morphologies.  Instead they tend to be more
"peculiar" in the sense that they
may represent mergers in progress \citep{abra96,bunk99}.  Although we detect changes in morphology between the two redshift
bins, these changes may not be solely attributed to evolution.  In Section 6.0 we discuss the possibility that the trends in morphology  may be
influenced by observational selection effects.

\section{Color Profiles}

We generated ($B_{450}-V_{606}$) and ($I_{814}-J_{110}$) color profiles 
for the 33 intermediate redshift and 50 high-redshift sample respectively and applied k-corrections 
in order to determine
if these galaxies exhibited significant color changes
(i.e. $\vert$ $\Delta(UV_{218}-U_{300})_o$/ $\Delta$log(r) $\vert$ $\geq$ 0.2 mag dex$^{-1}$) in their rest-frame $(UV_{218}-U_{300})_o$
profiles (Figures $6-8$).  
We can assume that 
since we are observing the color profiles for the rest-frame $(UV_{218}-U_{300})_o$, we will be
probing primarily the light from the young stars.  Thus, 
the sign of the gradient will give us an idea of the locations of the 
star forming regions if the color gradient is caused by an age gradient alone.  
If the galaxy possesses a \it positive\rm \ 
gradient, this would indicate a blueing towards the center, which might be 
explained by star formation that is centrally concentrated.  If the 
galaxy possesses a \it negative\rm \ gradient, this would point to a blueing with 
increasing radius which would imply an older stellar population in the center.  If, however, 
we observe no significant gradient in the galaxy, we can assume that the 
star formation regions are uniformly dispersed throughout the galaxy.  

Due to the $(1+z)^4$ fall off in the surface brightness, the color profiles
of our high redshift sample do not extend to the low surface brightness levels of our intermediate redshift sample.  For
example,
in our high redshift galaxies, we might reach depths of as much as 28 $mag/arcsec^2$, but due to the fall off in surface
brightness, this would translate to a depth of only $\sim$ 24.5 $mag/arcsec$ in our intermediate redshift sample.  In order
to
consistently compare the results for our two redshift bins, we must sample over approximately the same surface brightness
range,
thus, we must apply a cutoff in surface brightness in our intermediate
redshift sample.  Applying a cutoff in the radial range is approximately the same as applying the cutoff in surface
brightness.  Thus, we have chosen to fit the color profiles of the intermediate redshift
sample only out to a radius of 4 kpc.  This radius cutoff represents approximately twice the mean half-light radius of our high redshift
sample which
was determined by summing up all the light within approximately 28 $mag/arcsec^2$.  
Column (8) in Tables 3 and 4 lists the color gradients from the
weighted least-squares fit to the color profiles and their
associated errors for our intermediate redshift and high redshift samples, respectively.  The errors quoted include the
uncertainties from the fit and a conservative upper limit in the uncertainty from the systematic background subtraction ($\sim$
$\pm$ 0.037 mag/dex).   
  
From Table 3, column (8) we notice that 12 (36\%) of the galaxies in our intermediate redshift sample 
possess
positive gradients, but only two of them have gradients greater than 
0.2 mag dex$^{-1}$ which
indicates that they may have a strong central concentration of star formation perhaps indicative of
a burst. 
Although 64\% of the galaxies at intermediate redshifts possess
negative
gradients, only 39\% of them possess gradients less than -0.20 mag dex$^{-1}$.  Furthermore, the mean of the color gradients
of  
all the galaxies at intermediate redshifts is -0.091 $\pm$ 0.007 mag dex$^{-1}$ which is relatively shallow.  
Thus, although the negative color gradients indicates that, in general, the
colors of these galaxies are bluer in the outer parts of the galaxies,
since the mean of the gradient is not too steep, the color changes are gradual across the
galaxies.  
Column (8) in Table 4 lists the rest-frame $(UV_{218}-U_{300})_o$ color gradients and the associated errors for our high redshift sample determined
using all the
isophotes which reach up to depths $\sim$ 28 $mag/arcsec^2$. We notice
that the color gradients of the majority of the high-redshift
galaxies (72\%) are positive with a mean of 0.272 $\pm$ 0.007 mag dex$^{-1}$.  
Out of the 36 galaxies that have positive
gradients, 83\% of them have gradients greater than 0.200 mag dex$^{-1}$ indicating that there are
significant color changes across these galaxies.  

Figure 9 represents a plot of the redshift versus the color gradients of the galaxies in our two
redshift bins.   
When we compare the rest-frame $(UV_{218}-U_{300})_o$ color gradients for our intermediate redshift 
sample with our high redshift sample, we detect different trends in the two redshift bins.
In our intermediate redshift sample, we saw a slight trend for a blueing with increasing radius (negative color gradients), 
implying that the star formation regions may be located generally in the outer disks of the galaxies or that the older
stellar populations may be located towards the center.  
But given that the amplitude of the gradients were not too significant, we
conclude that in general the color changes are gradual across the galaxies.
However, in our high
redshift
sample, the majority of the galaxies possessed significant positive gradients in their color
profiles.  On average, these color gradients were 0.363 $\pm$ 0.010 mag dex$^{-1}$ higher 
than their lower redshift counterparts.  If these color gradients are due primarily 
to an age gradient, this would mean that
these galaxies have bluer colors
towards the center 
implying that their star formation regions 
may be centrally concentrated, i.e. there are more young stars at the centers 
than in the outer parts of the galaxies.  Numerous analysis show that both
present-day disks and spheroids have metallicity
gradients with higher metallicities toward the galaxy centers.  Such a metal gradient would only strengthen the 
age gradient needed to make the centers bluer.
Thus, from our analysis, we conclude that
there is evolution in the color profiles of galaxies in the HDF-N.  When we look further back
in time, the star formation regions of the galaxies are more centrally concentrated.  
Since it is beyond the scope
of this paper to generate models to interpret our results, in the next section, we discuss
modeling of color gradients taken from the literature in order to infer the dominant processes
responsible for producing the trends seen in the color gradients.  
 
\subsection{Discussion}

Not much is currently known about the ultraviolet color gradients of galaxies in the local universe.  
The few studies of this kind were conducted using the information from the images taken with the Ultraviolet Imaging
Telescope during the Astro-1 and 2 missions \citep{ocon92,ohl98}.  
From the Astro-1 mission \citet{ocon92} analized the (152-249) color profiles  of 
two Sb bulges and two E galaxies which were generated using the images taken with the 152 nm and 249 nm
filters.   They concluded that 
all but M32 exhibited prominent positive gradients in their profiles.  In their study of the FUV(1500\AA)-B color
profiles for eight early-type galaxies, \citet{ohl98} found, again, that all but M32 exhibited large positive
color gradients, with differences in color greater than 1.0 mag over the entire profile.  These large ultraviolet color
gradients are reflective of the "UV upturn" or "UVX" population which has been observed in the far-UV (1200 \AA $<$
$\lambda$ $<$ 2000 \AA) imaging of
local spiral bulges and early-type galaxies
(\citep{cod79,deB82}.  They
propose that a combination of an increase in metallicity combined with the existence of older stellar populations may contribute in
producing these color gradients.  The trends in the far-UV color gradients of these galaxies are in
the opposite sense to what we observe in the rest-frame near-UV for the majority of the intermediate redshift galaxies in
our HDF-N sample and local
galaxies which have been observed in the optical and near-IR.  However, since the study of the far-UV color
profiles of galaxies were conducted at shorter wavelengths, and given that the sample sizes were
much smaller and limited only to early-type systems and spiral bulges, it is difficult to compare it with our study.
Thus, more observations of local galaxies in the ultraviolet
are needed in order to fairly compare with our sample of galaxies at intermediate and high redshifts.     

In our intermediate redshift sample of galaxies, we find that most of the galaxies (6 out of 9) which we classified as having
$r^{1/4}$ profiles possess negative rest-frame $(UV_{218}-U_{300})$ color gradients indicating reddening towards the center of the
profile.  This is consistent with the rest-frame optical color profiles observed in the past.
In general early-type galaxies have been found to exhibit negative gradients, i.e. they have redder colors
in the center and gradually become bluer outwards \citep{tamu00,tam00,vad88,fran89,pel90}.  
In their study of the
($V_{606}-I_{814}$) color profiles of 10 elliptical
galaxies in the HDF-N from $z = 0.1 - 1.0$, \citet{tamu00}
observed the same
trends.   They found that all but two (one of them being galaxy 2-251 in our sample) exhibited redder colors 
towards the centers.          
At $z = 0$, both an age and a metallicity gradient in the stellar
populations can be responsible for producing the
negative color gradients observed in the galaxies, i.e. the galaxy colors are redder in the center than in the outer
parts because they either have older stellar populations  or higher metallicities there.  This age-metallicity
degeneracy makes it difficult to determine
the cause of the color gradient.  In order to 
break the degeneracy, they generated two model gradients which reproduce the typical color gradients of 
elliptical galaxies 
at $z = 0$, one
caused by a metallicity gradient of old stars, the other from an age
gradient of stars with the same metallicity.  They then evolved 
these models to the redshifts of their observed galaxy sample.  By comparing the model gradients with
the observed gradients, they concluded that
the model involving the metallicity gradient agreed well with observations at all
redshifts whereas the model involving the age gradient was only in good
agreement up to a redshift of z $\sim$ 0.3.  We also have ($V_{606}-I_{814}$) color profiles for the galaxies
in our intermediate redshift sample which reveal the same trends in the color gradients (Figure 10).  We again observe that 
over half (5 out of 9) of the galaxies possessing $r^{1/4}$ profiles exhibit negative color gradients.  Thus, we may interpret
the central reddening in the color
profiles of the elliptical galaxies in our intermediate redshift sample in the HDF-N as being 
due primarily to a metallicity gradient; the centers of
the galaxies have higher metallicities than in the outer parts.
We observe the same negative trend in the color gradients for the majority of the intermediate redshift 
galaxies (14 out of 23) which 
we classified as peculiars/irregulars or having exponential profiles.  
In the study of the near-IR and optical color profiles of 86 face-on disk
dominated galaxies, \citet{deJ96} concluded that their negative color gradients were
best reproduced by models involving both stellar age and metallicity gradients. 
Therefore, we interpret the gradients seen in the late-type galaxies as arising as a result of 
both age and metallicity effects; the central parts
of the galaxies in general have older stars and higher metallicity than the
outer parts making them have redder colors towards the centers.

Whereas the trends seen in the color profiles of our intermediate redshift sample are well understood, 
we can not say the same for our high redshift sample. 
Since this is the first study of its kind to be conducted on galaxies with z $\geq$ 2.0,
there are no other high redshift observations of color profiles to compare our results with.  Thus,  
we must attempt to account for the positive gradients
seen
in our high redshift sample by 
understanding what is responsible for producing similar trends in the color profiles seen at intermediate and low redshifts.
\citet{guz98} generated
rest-frame $(B-V)$ color profiles for five compact narrow emission line galaxies
(CNELGs) which are low mass ($M \leq 10^{10}$
$M_{\odot}$) starburst systems seen at $z \leq 1.0$ and are similar to nearby HII
galaxies.  They noticed that although the gradients
were not significant, there was a slight trend for the bluest color to occur in
the central regions.  On average, they found that the color inside the half-light radius is
$0.14 \pm 0.05$ mag
bluer than outside the half-light radius.  \citet{guz98} futhermore
describes the CNELGs as having either compact cores, fans and tails, or several
small compact regions.  These morphologies are along the lines of what we see in
our high redshift sample.  They also determined half-light radii to span from $1-5$
kpc ($H_o=50 km/s/Mpc$, $q_o=0.1$) which match the range we see in our high redshift
sample.   Given that the morphologies, half-light radii, 
and color profiles of the CNELGs are similar to what we observe in the high redshift
sample, it is not unreasonable to believe that CNELGs could be the low redshift, lower-luminosity
analogs of the higher redshift Lyman break galaxies.   We can then interpret the positive gradients seen in our
high redshift sample
as being caused by centrally concentrated starbursts.   

The high redshift galaxies have been shown to possess modest to high star formation rates ($4-25$
$h_{50}^{-1}$ $M_{\odot}$/yr, $q_{o}$=0.5) \citep{low97,steid95,pet98} which further enforces the idea
that the positive gradients seen in our sample could be caused by centrally concentrated starbursts.
Starbursts can be triggered by several mechanisms including mergers and interactions \citep{mh94,mh95,mh96,bar96,SPF00}, bar
instabilities \citep{shlos90}, and SNe and stellar winds, e.g. \citet{heck90}.  
Since the volume of the high redshift universe is smaller than it is now, mergers have been implied to be the
cause of the enhanced star formation, however, this topic is still under debate.  
In Section 7, we
argue that the starbursts which may be responsible for causing the central blue colors of our high redshift sample 
are most likely ignited through mergers and interactions.  

\subsection{Sources of Uncertainties}

When we discuss the trends seen in the color profiles, we must take into account the uncertainities involved. These
uncertainties include
the determination of the sky background, the correlation of the pixels,  and the role of dust.
While statistical
errors dominating the inner parts of the color profiles and PSF effects contribute to the uncertainty in the least-squares
fit, the systematic errors due to sky subtraction dominate in the outer parts of the profile .  In order to assess how our
color gradients are affected
by sky subtraction, we chose 20 high redshift galaxies possessing positive color gradients and for each
we obtained statistics on 11x11 pixel wide grids in six blank areas of the sky around the object in the $I_{814}$ and the
$J_{110}$ passbands. 
We generated a histogram of the 726 total pixels in the six grids to determine the mean and the standard deviation
of the mean ($\sigma$).  For each galaxy, we then determined both a lower and an upper limit of the sky values in $I_{814}$ and
$J_{110}$  by subtracting
3$\sigma$ from the means to obtain their lower limits (I-, J-) and adding 3$\sigma$ to the means to obtain their upper limits
(I+,J+). Lastly, we generated their ($I_{814}-J_{110}$) color profiles and obtained their color gradients for each of
the five
possible
combinations of sky values subtracted off ((1) mean in $I_{814}$, mean in $J_{110}$ (2) I- ,J- (3) I+,J+ (4) I-,J+ (5) I+, J-).  
The mean in the color gradients for the five different cases respectively are 0.3722, 0.425, 0.321, 0.249, and
0.503 which represent a 3$\sigma$ deviation of $\sim$ 0.112 mag/dex making the typical uncertainty in the estimation of
the sky
background to be $\sim$ 0.037 mag/dex.
As we can see, the mean value of the color gradients remain positive and greater than 0.200 mag/dex in all cases.  
Thus, we find that the typical gradients seen in the high
redshift 
galaxies ($\sim$ 0.26 mag/dex) are significant enough, that even with the uncertainties involved in the estimation of the sky
background, 
our main conclusion that these galaxies have centrally concentrated star formation still holds.

Another potential source of concern is the correlation of the pixels.  Since the images have been drizzled, convolved to recover approximately the
same resolution as the
$H_{160}$ image, and fitted with ellipses at 0.1 pixel increments, the errors that we obtain from the photometry will be correlated due to the
oversampled pixels.  Thus, when we performed a weighted least-squares fit to the color profiles, the error bars used in
the
determination of the weights were
correlated.  Consequently, the parameters that we obtain from the fit, i.e. the color gradient in this case,
may be off from what would be measured if the errors had been independent.  To understand how the correlated errors in our photometry affect the
least-squares fitting to a straight line, we obtained rms images 
(M. Dickinson 2001, private communication)  
in the passbands of interest.  These images quantify the
uncertainty of the noise in the background (sky + readout noise + dark current) before any processes were performed to cause interpixel correlations.
We re-analyzed ten random galaxies from the intermediate and high redshift sample using the rms images.  With
the same
set of ellipses that we used in the surface photometry, we replaced the correlated errors with the "true"
uncorrelated errors for each galaxy.  In doing so, we
have chosen to neglect the poisson noise from the galaxies themselves since they are so faint that it would
be insignificant compared to the background noise.   From this analysis, we find that even after taking into
account interpixel correlations, the sign of the color gradients for the galaxies does not change.  The high redshift
galaxies still
possess
predominantly
positive color gradients whereas the intermediate redshift galaxies possess predominantly negative color gradients.
         
However, the rest-frame $(UV_{218}-U_{300})_o$ color profiles may be affected by the presence of dust.  
In theory, dust can cause the
color profiles of galaxies to be
redder towards the center if we assume that the dust would be more concentrated at
the center and  would produce more extinction there \citep{deJ96,evans94,buy94}.  \citet{deJ96} conducted Monte Carlo
simulations of light rays traveling through a dusty medium in order to
understand to what extent dust can affect color gradients in late-type systems. 
They concluded that
although a fraction of the color gradients could be attributed to dust, an 
additional explanation is needed to account for the total reddening in the
system.  \citet{wise96} produced color profiles extincted by dust
for a set of model ellipticals from the U to the K passband and compared
them with 52 early-type galaxies from the literature.  They concluded that if
dust is the primary cause of color gradients then the dust must be spatially
extended ($\rho_d$ $\propto r^{-1}$) implying high dust masses comparable to the
dust masses inferred by the IRAS data.   Thus, the presence of dust may be partly responsible
for producing the color gradients seen in our intermediate redshift sample for both early-type and late-type systems, 
however, given that the total mass of the dust and its spatial distribution are not well constrained, 
we can not say, for now, to what degree dust affects the color
distribution in the galaxies.  On the other hand, even though dust might play a
role in our high redshift sample, 
there is no reason
why its presence would cause the color profile to be bluer towards the center.  As previously
mentioned, dust would tend to cause extinction and, thus, reddening towards 
the center, not the outer
parts of the galaxies.   In fact, if dust is found to cause extinction in the centers of these galaxies, then 
the color gradients seen in our high redshift sample would be even greater.  Thus,
although dust may play a role in generating the
color gradients of our intermediate redshift sample, it can not explain the trends seen
in our high redshift sample.  

\section{Selection Effects}

Anytime we compare observations of galaxies in two different redshift 
ranges, we must take into account the effect of observational selection.  At high redshifts, we will be able to detect and study only 
those galaxies with the highest luminosities and surface brightnesses, thus, the trends that we see in the surface brightness and
color profiles may potentially be affected by this bias.  
We have tried to minimize some selection effects by observing the galaxies at different redshifts in the same rest-frame
wavelengths.  This guarantees that at all redshifts, we will be probing the same population of stars.  Furthermore, at high redshifts
we sample not only galaxies with spectroscopic redshifts, but also galaxies with photometric redshifts which are, in general, fainter
than  those galaxies that can be detected spectroscopically.  
Thus, we are sampling the fainter population at high redshifts and are
not only just picking out the UV luminous galaxies.  

\citet{wu99} performed simulations on galaxies in the HDF-N in order to 
understand the selection effects that may arise when studying galaxies at different redshifts.  
She takes the galaxies with $0.75 \le z \le 1.2$ in the F814W band and 
simulates their appearance at $z = 1.5$ and $z = 2.3$ by using the corresponding images in F606W and
F450W filters respectively to minimize k-corrections.  She finds that at $z = 1.5$ only 33\% of the original objects can still be
detected and by $z = 2.3$, only 3\% of them remain.   To further
understand the
sampling of the galaxy population as a function of redshift, she defines a
 concentration versus asymmetry index ($A_W - C_W$). She plots the $A_W-C_W$ relation for galaxies in the HDF
separating them into five redshift bins with redshifts ranging from
z $\leq$ 0.6 to redshifts z $>$ 3.0.  From these plots, she concludes that whereas the range of $A_W$ and $C_W$ values remain constant for z $<$
1.2, at higher redshifts, the number of high $C_W$ objects seem to decrease while the number of high $A_W$  objects seem to increase.
Since high concentration is associated with early-type systems such as ellipticals and early-type spirals and high asymmetry is
associated with merging systems, it seems from her plots that at z $>$ 1.2, there is a decrease in the early-type systems and
an increase in merging systems.  
However, she shows in her work, that these trends can naturally be explained by selection effects.  In her simulations she has
shown that, by $z = 2.3$ most of these galaxies with high concentration have faded so much
that they "disappear" into the noise and can no longer be detected.  

From Wu's (1999) discussion it is quite possible that the trends we see in the surface
brightness profiles may be influenced by these selection effects.  
In our work with the surface brightness profiles, if we had not corrected for the PSF effect, we would have seen the same
decrease
in early-type systems.  
If we just inspect the S\'ersic indices of the galaxies and ignore the effect of PSF smearing, we would naively classify  5 out of
a total of 32 (16 $\pm$ 7\%) of the galaxies in the intermediate sample and only 1 out of the 20 (5 $\pm$ 5\%) galaxies with H $\leq$
24.5 in the high redshift sample as possessing an $r^{1/4}$ profile.  Thus, it would seem that the number of
early-type 
systems had decreased with redshift.  However, since we have corrected for the effect of the PSF, we recover the number of galaxies with $r^{1/4}$
profiles (i.e. 28 $\pm$ 9\% for the intermediate redshift sample and 60 $\pm$  15\% in the high redshift
sample).  Thus, by correcting for the
PSF, we take care
of one selection effect.  
However at z $>$ 2, we are still sampling
only the tip of the surface brightness function where asymmetries are more evident. 
Consequently, it is not surprising that we would find a slightly higher
fraction of "irregular"
galaxies in the high redshift sample, 20 $\pm$ 10\% as opposed to 6 $\pm$ 3\% in the intermediate redshift
sample.    
Although the selection effects may be responsible for some of the trends seen in the morphological classification of galaxies, they cannot
explain the trends that exist in the color profiles.  \it They cannot explain why at high
redshifts, we detect a significant population of
galaxies possessing large positive color gradients but at lower redshifts, they seem to
"disappear" \rm.
	                
\section{Comparison with Theoretical Models}

In order to fully comprehend galaxy formation and evolution, it is 
important to compare observations with theoretical models.  However,  
only in the last decade has it been 
possible to study 
galaxy formation observationally.  Previously, we were only able to 
predict how galaxies form and evolve by using N-body simulations 
and semi-analytical modeling  of 
dark matter halos, which demonstrated 
that structure builds up into larger and larger units through continual 
accretion and merger \citep{bau98,cole94,kauf97,rouk97,whit91}.  These models gave support to the "hierarchical 
clustering" scenario pioneered by \citet{peeb70} which is a less restricted version of the standard
cold dark matter (CDM) model since it allows for the assumption of different 
cosmologies.  In the last several years, much advancement has been made
in the observational regime since many have used the Lyman break 
technique \citep{steid93} to 
reliably detect galaxies at high redshift in large numbers.   With the
detection of numerous high redshift 
galaxies at high resolution and sensitivities, it is now feasible to  
compare observations with theoretical predictions.  

From our surface brightness analysis of the galaxies in the HDF-N, we find 
that as we look further back in time, the star formation regions are more centrally concentrated and 
their morphologies are more irregular, deviating from the traditional Hubble sequence.
Although our results may be affected by uncertainties and selection effects, we
can compare to models to test for consistency and determine what may be the driving force behind the trends seen if they are at the
very least
attributed in some way to galaxy evolution.  
We have discussed the possibility of mergers and interactions as the cause of the trends seen.  
In order to understand how our results fit in with the overall picture of galaxy formation and evolution,
we compare our observations of
galaxies in the HDF-N with the predictions 
from the semianalytic model of galaxy formation described by BCFL and SPF 
which focuses particularly on 
the high redshift Lyman break galaxies.  

In their model, BCFL used a Monte Carlo 
simulation to generate binary merger trees which essentially describes 
the merging histories for dark matter halos with predetermined final 
masses.   They generated the mock high redshift 
progenitors by applying the selection criteria for Lyman break 
candidates described in \citet{steid93} and quantified the 
amount of 
cold gas available by measuring the rate at 
which cold gas in thin shells can cool to form stars.  They chose values 
for the star formation timescale, the feedback parameter, the shape of the 
initial mass function (IMF), the overall luminosity normalization, and 
the merger timescale for galaxies that would best reproduce the local B 
and K band luminosity functions.  Lastly, they fixed the threshold mass 
for a galaxy merger causing a disk to become a spheroid to match 
the relative population of ellipticals, S0's, and spirals locally.
They produced their model for three different cosmologies: the standard 
CDM model ($\Omega_0=1$, $h=0.5$, $\sigma_8=0.67$), the flat CDM model  
($\Omega_0=0.3$, $\Lambda_0=0.7$, $h=0.6$, $\sigma_8=0.97$), and the open 
CDM model ($\Omega_0=0.4$, $h=0.6$, $\sigma_8=0.68$).  

From their simulations, 
BCFL show that present day galaxies that make up the 
bright end of the luminosity function 
have at least one Lyman break progenitor at $z \sim 3$.  Furthermore, 
depending on their merging histories, these Lyman break progenitors 
can end up as a L $>$ L* galaxy with any morphological type. If they 
experienced major mergers at recent epochs they may evolve to become 
ellipticals and S0s. If they grew by the quiescent accretion of 
cooling gas, the gas might form a disk around a bulge that was 
created by an earlier merger event. And if they experienced minor 
mergers which are too weak to destroy the disk, they might 
accumulate more stars in the central bulge.  They expect the Lyman break 
galaxies to be rotating disks with typical half-light radii 
$\sim 0.5h^{-1}kpc$ and predict 
that the stars of the progenitors will congregate primarily in 
the central regions.  

BCFL additionally conclude that the star formation of galaxies at all redshifts
occurs quiescently with SFRs that never obtain a value higher than a few solar masses per year.
Their conclusion matches the observations of LBGs only if internal dust extinction
is ignored.  However, in reality, the effects of dust is probably
non-negligible although how much of a role 
it plays is still presently uncertain. \citet{pet97} measured the
extinction to be a factor of $\sim$ 3 at 1500\AA \ whereas \citet{meurer97} and \citet{sy98}
 report values closer to $15 - 20$.  Now the general consensus is to use an extinction correction of $\sim$ 5 \citep{mhc99,steid98}.   
If the central blueing seen in the color profiles of our high redshift sample
is caused by a central starburst, which seems likely, then BCFL's assumption of quiescent star formation in an extended
 disk does not agree with 
our observations.  On the other hand, the results of the model from SPF would agree well with what we observe
in the color profiles and with what we
currently know about the SFRs of LBGs.  Instead of undergoing quiescent star formation, they favor the
idea that LBGs are experiencing what they 
term as "collisional starbursts" which describe starbursts that are induced by galaxy-galaxy mergers.      

SPF explain that they obtain different results from BCFL because of the different assumptions made in the two
models.  The assumptions made in BCFL for supernovae feedback, gas cooling, and types of mergers were such 
that, taken in combination, would act to suppress star formation at early epochs (please refer to BCFL and SPF
for the specifics on the differences in the two models).  SPF use a semi-analytic modeling of galaxies in 
the CDM hierarchical clustering framework adopting the $\Lambda$ CDM cosmology ($\Omega_0 = 0.3$, $\Omega_{\Lambda} = 0.7$,
$h=H_0/(100km/s/Mpcs) = 0.7$, $\sigma_8 = 1.0$)  to understand the
nature of Lyman break galaxies. Their models take 
into account
the effects of the cooling rate of gas, supernovae feedback,  galaxy-galaxy mergers, and star formation and
subsequent evolution.   They address the effects of dust extinction by following the recipe described in Wang \& Heckman (1996)
for nearby starbursts which returns the optical depth for a given extinguished UV luminosity, intrinsic UV luminosity, and
face-on optical depth.   They
generate
Monte-Carlo based merger trees in which each branch in the tree is associated
with a merger.
Within this framework, they considered three different models to represent the variety of star formation events.  
The "collisional starburst" model is one in which star formation occurs quiescently with 
additional star formation occurring in bursts.  The "constant efficiency quiescent" model only involves 
quiescent star formation with no bursts. And the "accelerated quiescent" model involves quiescent star formation 
but with the star formation rate increasing with look-back time.  
SPF conclude that among the three models, the "collisional starburst" model best reproduces the observations of the 
LBGs including comoving number density, star formation rates, internal velocity dispersions and the observed luminosity function, global star
formation rate density, and cold gas and metallicity abundances in the Universe.  
Thus, they advocate that star formation occurred at a higher rate in the high redshift universe and 
that mergers were responsible for increasing the 
star formation rate by triggering gas inflow to the central regions.  
  
The positive color gradients we see at high redshifts would then be associated with nuclear starbursts
induced by mergers and/or interactions, which could also cause star formation to "migrate"
over time from the central regions to the spiral arms where we see most of the star
formation today.  That is, when galaxies are first formed, the stars would tend to be born in
the dense central regions.  Then some galaxies may undergo mergers, which 
funnel gas to the central regions
causing a burst of star formation.  After awhile,  the gas will be depleted and dispersed 
such that during the late stages of merger, the central light profiles will evolve to
follow the $r^{1/4}$ law \citep{mih92,neg83,bar92}.  During the late stages of the merger, the presence of a central bulge might suppress 
the inflow of gas and consequently reduce the star formation efficiency \citep{mh94}. 
Then if the galaxy experiences further interactions with another galaxy, material
can rain in on the outer parts causing 
the gas to collapse to form stars around the spheroid.  
Therefore, if the hierarchical structure formation scenario represents an accurate prediction of 
structure formation,
then at redshifts between 2.0 $\leq$ z $\leq$ 3.5, 
we may be witnessing the epoch at which galaxies are experiencing mergers 
 and their star formation is enhanced and centrally concentrated. 

\section{Conclusions}
From our surface brightness and color analysis of galaxies at a range of
redshifts probed at the same rest-frame wavelengths in the Hubble Deep Field-North, we conclude:

1). The color profiles of galaxies reveal that
at earlier epochs, there are fewer galaxies with old red stellar
populations in their centers and most galaxies have centrally concentrated star
formation.  As in local samples of galaxies, our intermediate redshift galaxies (0.5 $\leq$ z $\leq$ 1.2) have redder central 
regions due to a combination of age, metallicity and dust
gradients.  Since age gradients are the only viable explanation for the central blueing of our high
redshift sample, we
conclude that the majority of high redshift galaxies contain centrally concentrated starbursts.  

2). From the galaxy models, we have demonstrated the importance of taking into account the PSF effects before
attempting to classify morphologies. 
After correcting for the PSF effects in the surface brightness profiles, we find that the
population of galaxies possessing $r^{1/4}$ profiles have slightly increased with look-back time 
while those with exponential profiles have 
slightly decreased.  Our results also suggest a statistically insignificant increase in the fraction of galaxies with peculiar/irregular
structure at higher redshifts. The
results should be viewed with a note
of caution, however, 
due to the small sample size and large uncertainties in the galaxy models especially at small scale lengths and faint magnitudes.    

3). If, as predicted by hierarchical galaxy formation models,  mergers and interactions played an important role in the
lifetime of these galaxies, such processes would be responsible for the positive color gradients indicative of centrally concentrated 
star formation seen in the high redshift galaxies. 
SPF show that their "collisional starburst" model, in which bursts are triggered 
by mergers, best reproduces the
observed properties of Lyman break galaxies and of the Universe in general.    
Such centrally condensed nuclear star bursts are consistent with our observed trends in the color gradients. 
A good way to test whether the "collisional starburst" model holds is by applying the color-asymmetry diagram developed by 
\citet{con00} to distinguish between starbursts driven by mergers and interactions from
starbursts which were ignited by some other method.

Since the Hubble Deep Field  represents only a small volume of space, we must be
cautious in generalizing our results to the universe as a whole.  We need to obtain more observations
of high redshift galaxies to measure critical quantities such as their virial masses and estimation of dust content.  
We also need to better understand how and to what degree observational
selection comes into play when we compare galaxies at different redshifts.  Given that we are now detecting fainter galaxies
at higher redshifts and that instruments with multi-object spectroscopic capabilities in the near-IR and 8-m class telescopes
are now readily available, much
information will undoubtedly be unraveled in the near future concerning the nature of LBGs which will allow us to formulate
a more robust theory on the formation and evolution of galaxies.

\acknowledgments
We would like to thank R. Guzm\'an for his much
appreciated suggestions during the finishing stages of the paper, M. Dickinson for
providing
extremely informative answers to questions about the data, K. Wu for her
support and useful discussion on selection effects, Devriendt et al. for
kindly emailing us their SEDs,  R. Pina and J.
Radomski for their help on IDL,
and E. McKenzie for a useful discussion on K-corrections.  Funds for this research were made possible by 
the CAREER Grant AST-9875448
and NASA grant "NICMOS MAP of the HDF" GO-07817.05-96A.  P. Moth also
acknowledges funding from the Florida Space Grant Fellowship.

\clearpage
\input{Moth.tablist.tex.rev}

\input{Moth.figcap.tex.rev}

\end{document}

%% file: Moth.tablist.tex.rev
\clearpage
\input{Moth.tab1.tex.rev}

\clearpage
\input{Moth.tab2.tex.rev}

\clearpage
\input{Moth.tab3.tex.rev}

\clearpage
\input{Moth.tab4.tex.rev}

\clearpage
\input{Moth.tab5.tex.rev}
\clearpage

%% file: Moth.tab1.tex.rev
\begin{deluxetable}{cccccccccc}
\tabletypesize{\small}
\tablecaption{ Half-light Radii and S\'ersic Indices of Convolved Model Disks ($n = 1$) and Spheroids ($n = 4$) in the $I_{814}$
Band }
\tablewidth{0pt}
\tablehead{ID  & Profile  & $R_{21}$\tablenotemark{a} & $n_{21}$\tablenotemark{b} & $R_{22}$\tablenotemark{c} &
$n_{22}$\tablenotemark{d} & $R_{23}$\tablenotemark{e} & $n_{23}$\tablenotemark{f} & $R_{24}$\tablenotemark{g}
& $n_{24}$\tablenotemark{h}}
\startdata

Gal1 & Disk & 1.012  & 0.781  & 0.984 & 0.790 & 0.928 & 0.771 & 0.840 & 0.805\\
Gal2 & Disk & 0.956 & 0.816  & 0.956 & 0.800 & 1.024 & 0.789 & 0.900 & 0.630\\
Gal3 & Disk & 0.844 & 0.789   & 0.816 & 0.750 & 0.772 & 0.689 & 0.696 & 0.582\\
Gal4 & Disk & 0.800 & 0.793   & 0.800 & 0.769 & 0.836 & 0.746 & 0.768 & 0.676\\
Gal5 & Disk & 0.712 & 0.785  & 0.712 & 0.765 & 0.756 & 0.747 & 0.720 & 0.732\\
Gal6 & Disk & 0.628 & 0.773  & 0.628 & 0.744 & 0.596 & 0.714 & 0.600 & 0.681 \\
Gal7 & Disk & 0.556 & 0.740  & 0.552 & 0.713 & 0.536 & 0.690 & 0.508 & 0.690\\
Gal8 & Disk & 0.480 & 0.727  & 0.472 & 0.696 & 0.460 & 0.671 & 0.440 & 0.643\\
Gal9 & Disk & 0.408 & 0.692  & 0.400 & 0.678 & 0.392 & 0.655 & 0.380 & 0.631\\
Gal10 & Disk & 0.308 & 0.707 & 0.304 & 0.694 & 0.300 & 0.668 & 0.296 & 0.630\\
Gal11 & Spheroid & 0.804 & 1.651 & 0.764 & 1.366 & 0.704 & 1.094 & 0.604 & 0.910\\
Gal12 & Spheroid & 0.832 & 1.736 & 0.804 & 1.435 & 0.676 & 1.209 & 0.568 & 1.013\\
Gal13 & Spheroid & 0.748 & 1.603 & 0.692 & 1.336 & 0.596 & 1.127 & 0.624 & 0.987\\
Gal14 & Spheroid & 0.704 & 1.572 & 0.656 & 1.360 & 0.596 & 1.151 & 0.440 & 0.910\\
Gal15 & Spheroid & 0.604 & 1.521 & 0.560 & 1.337 & 0.496 & 1.111 & 0.432 & 0.906\\
Gal16 & Spheroid & 0.564 & 1.517 & 0.548 & 1.346 & 0.508 & 1.115 & 0.452 & 0.914\\
Gal17 & Spheroid & 0.508 & 1.480 & 0.504 & 1.325 & 0.492 & 1.114 & 0.456 & 0.913\\
Gal18 & Spheroid & 0.436 & 1.368 & 0.428 & 1.266 & 0.384 & 1.089 & 0.352 & 0.917\\
Gal19 & Spheroid & 0.388 & 1.320 & 0.392 & 1.227 & 0.388 & 1.108 & 0.400 & 0.984\\
Gal20 & Spheroid & 0.300 & 1.088 & 0.300 & 1.063 & 0.296 & 0.992 & 0.304 & 0.870\\
\enddata
\tablenotetext{a}{ Half-light radii in arcsec for $I_{814} = 21$}
\tablenotetext{b}{ S\'ersic index for $I_{814} = 21$}
\tablenotetext{c}{ Half-light radii in arcsec for $I_{814} = 22$}
\tablenotetext{d}{ S\'ersic index for $I_{814} = 22$}
\tablenotetext{e}{ Half-light radii in arcsec for $I_{814} = 23$}
\tablenotetext{f}{ S\'ersic index for $I_{814} = 23$}
\tablenotetext{g}{ Half-light radii in arcsec for $I_{814} = 24$}
\tablenotetext{h}{ S\'ersic index for $I_{814} = 24$}
\end{deluxetable}

%% file: Moth.tab2.tex.rev
\begin{deluxetable}{cccccc}
\tabletypesize{\small}
\tablecaption{Half-light Radii and S\'ersic Indices of Convolved Model Disks ($n = 1$) and Spheroids ($n = 4$) in the $H_{160}$ Band }
\tablewidth{0pt}
\tablehead{ID  & Profile  & $R_{23}$\tablenotemark{a} & $n_{23}$\tablenotemark{b}
& $R_{24}$\tablenotemark{c} &
$n_{24}$\tablenotemark{d} }
\startdata

Gal1 & Disk & 0.488  & 0.754  & 0.288 & 0.605 \\
Gal2 & Disk & 0.432 & 0.672   & 0.384 & 0.623 \\
Gal3 & Disk & 0.372 & 0.635   & 0.364 & 0.630 \\
Gal4 & Disk & 0.292 & 0.595   & 0.300 & 0.591 \\
Gal5 & Disk & 0.252 & 0.593  & 0.248 & 0.582 \\
Gal6 & Disk & 0.232 & 0.590  & 0.228 & 0.554 \\
Gal7 & Disk & 0.232 & 0.622  & 0.236 & 0.588 \\
Gal8 & Disk & 0.220 & 0.618  & 0.208 & 0.599 \\
Gal9 & Disk & 0.200 & 0.615  & 0.208 & 0.609 \\
Gal10 & Disk & 0.188 & 0.622 & 0.196 & 0.592 \\
Gal11 & Spheroid & 0.416 & 0.848 & 0.356 & 0.716 \\
Gal12 & Spheroid & 0.392 & 0.892 & 0.368 & 0.804 \\
Gal13 & Spheroid & 0.364 & 0.904 & 0.332 & 0.771 \\
Gal14 & Spheroid & 0.296 & 0.794 & 0.304 & 0.783 \\
Gal15 & Spheroid & 0.272 & 0.785 & 0.260 & 0.715 \\
Gal16 & Spheroid & 0.248 & 0.760 & 0.240 & 0.731 \\
Gal17 & Spheroid & 0.240 & 0.744 & 0.236 & 0.690 \\
Gal18 & Spheroid & 0.224 & 0.715 & 0.240 & 0.686 \\
Gal19 & Spheroid & 0.208 & 0.690 & 0.196 & 0.636 \\
Gal20 & Spheroid & 0.204 & 0.687 & 0.212 & 0.681 \\
\enddata
\tablenotetext{a}{ Half-light radii in arcsec for $H_{160} = 23$}
\tablenotetext{b}{ S\'ersic index for $H_{160} = 23$}
\tablenotetext{c}{ Half-light radii in arcsec for $H_{160} = 24$}
\tablenotetext{d}{ S\'ersic index for $H_{160} = 24$}
\end{deluxetable}

%% file: Moth.tab3.tex.rev
\begin{deluxetable}{cccccccc}
\tabletypesize{\small}
\tablecaption{0.5 $\leq$ z $\leq$ 1.2 Galaxies}
\tablewidth{0pt}
\tablehead{Id\tablenotemark{a} & z\tablenotemark{b}  &
$M_I$\tablenotemark{c} & n($I_{814}$)\tablenotemark{d} & Class\tablenotemark{e} &
R($I_{814}$)& R($I_{814}$)&   
$\Delta(UV_{218}-U_{300})_o$/$\Delta$log(r)\tablenotemark{f} \\
 & &  & & & (arcsec) & (kpc) &   
}

\startdata
2-353 & 0.609 & 23.67 & 0.6908 $\pm$ 0.0008 & Exp & 0.328 & 1.867 & 0.1873 $\pm$ 0.0376\\
2-251 & 0.962 & 21.32 & 2.4128 $\pm$ 0.0020 & deV & 0.472 & 3.056 & 0.8181 $\pm$ 0.0402\\
2-661 & 0.819 & 23.22 & 0.7738 $\pm$ 0.0004 & Exp & 0.732 & 4.574 & -0.1234 $\pm$ 0.0375\\
2-514 & 0.752 & 21.84 & 0.9400 $\pm$ 0.0003 & Int & 0.444 & 2.711 & -0.0359 $\pm$ 0.0384\\
2-950 & 0.517 & 23.60 & 0.6345 $\pm$ 0.0006 & Exp & 0.564 & 3.004 & -0.0858 $\pm$ 0.0371\\
2-809 & 0.498 & 23.44 & -   & O & 0.452 & 2.368 & -0.2876 $\pm$ 0.0464\\
2-702 & 0.559 & 23.09 & 0.7001 $\pm$ 0.0012 & Exp & 0.544 & 2.995 & 0.0381 $\pm$ 0.0373\\
2-652 & 0.557 & 21.09 & 0.8540 $\pm$ 0.0004 & Exp & 0.676 & 3.716 & 0.3950 $\pm$ 0.0374\\
2-1023 & 0.564 & 23.02 &0.6237 $\pm$ 0.0005 & Exp & 0.744 & 4.110 & -0.0446 $\pm$
0.0372\\
1-105 & 0.846 & 24.04 & 0.8923 $\pm$ 0.0014 & deV & 0.264 & 1.663 & -0.6397 $\pm$ 0.0538\\
2-1018 & 0.559 & 23.75 & 1.0071 $\pm$ 0.0011 & deV & 0.620 & 3.413 & -0.1995 $\pm$
0.0374\\
2-860 & 0.851 & 22.56 & 0.6221 $\pm$ 0.0002 & Exp & 0.540 & 3.406 & -0.3035 $\pm$ 0.0372\\
2-982 & 1.148 & 22.81 & 0.8546 $\pm$ 0.0015 & Int & 0.784 & 5.210 & -0.2074 $\pm$ 0.0386\\
2-402 & 0.557 & 20.97 & 0.8346 $\pm$ 0.0005 & Exp & 0.824 & 4.530 & -0.1373 $\pm$ 0.0374\\
4-493 & 0.849 & 21.70 & 1.9410 $\pm$ 0.0014 & deV & 0.436 & 2.749 & -0.4363 $\pm$ 0.0373\\
4-89  & 0.681 & 24.04 & 0.6743 $\pm$ 0.0008 & Exp & 0.520 & 3.080 & 0.0982 $\pm$ 0.0373\\
4-132 & 0.511 & 24.10 & 0.5805 $\pm$ 0.0009 & Exp & 0.416 & 2.204 & 0.1916 $\pm$ 0.0421\\
3-430 & 0.875 & 23.52 & 0.6693 $\pm$ 0.0016 & Exp & 0.612 & 3.886 & -0.2257 $\pm$ 0.0373\\
4-950 & 0.609 & 22.27 & 0.8792 $\pm$ 0.0007 & Int & 0.620 & 3.530 & -0.1245 $\pm$ 0.0384\\
4-221 & 0.952 & 22.94 & 0.7173 $\pm$ 0.0012 & Exp & 0.599 & 3.877 & 0.1406 $\pm$ 0.0371\\
3-350 & 0.642 & 20.97 & 0.8820 $\pm$ 0.0005 & Exp & 0.728 & 4.224 & -0.4566 $\pm$ 0.0373\\
3-404 & 0.520 & 23.58 & 0.7805 $\pm$ 0.0011 & Exp & 0.472 & 2.520 & 0.1730 $\pm$ 0.0377\\
4-928  & 1.015 & 22.32 & 1.4232 $\pm$ 0.0010 & deV & 0.356 & 2.327 & -0.3160 $\pm$ 0.0420\\
3-610 & 0.518 & 19.95 &  2.3243 $\pm$ 0.0152     & deV,Exp &1.496 & 7.975 & -0.4149 $\pm$ 0.0375\\
      &       &       &  0.8263 $\pm$ 0.0020  &     &      &       &    \\
3-551 & 0.560 & 22.95 & 0.5967 $\pm$ 0.0004 & Exp & 0.544 & 2.997 & 0.0954 $\pm$
0.0374\\
3-378 & 1.225 & 24.06 & - & O & 0.8320 & 5.564 & -0.2685 $\pm$ 0.0373\\
4-565 & 0.752 & 22.74 & 1.1104 $\pm$ 0.0014& deV & 0.352 & 2.149 & 0.0763 $\pm$ 0.0372\\
3-486 & 0.790 & 21.99 & 1.1928 $\pm$ 0.0009 & deV & 0.560 & 3.467 & -0.6729 $\pm$ 0.0372\\
4-727 & 1.242 & 23.18 & 1.0207 $\pm$ 0.0012& deV & 0.316 & 2.116 & 0.0849 $\pm$ 0.0371\\
4-775 & 1.010 & 22.50 & 0.9882 $\pm$ 0.0009 & Int &  0.632 & 4.126 & -0.3229 $\pm$ 0.0383\\
4-850 & 0.961 & 24.61 & 0.6211 $\pm$ 0.0013& Exp & 0.380 & 2.460 & 0.0983 $\pm$ 0.0375\\
4-260 & 0.962 & 22.88 & 0.8264 $\pm$ 0.0013& Exp & 1.004 & 6.500 & -0.2918 $\pm$ 0.0371\\
3-321 & 0.680 & 21.43 & 1.904 $\pm$ 0.0027& deV & 0.560 & 3.311  & -0.0350 $\pm$ 0.0377\\
\enddata
\tablenotetext{a}{ Ids are from Williams et al. 1996}
\tablenotetext{b}{ Spectroscopic redshifts are from Cohen et al. 1996 and Zepf et al. 1997}
\tablenotetext{c}{ F814W AB mag within $\sim$ $26 \ mag/arcsec^2$}
\tablenotetext{d}{ S\'ersic index in rest-frame B}
\tablenotetext{e}{ deV = de Vaucouleur, Exp = Exponential, Int = Intermediate, O = Other}
\tablenotetext{f}{ Rest-frame $(UV_{218}-U_{300})_o$ color gradient within $R_{1/2}$ $\sim 4 \ kpc$}
\end{deluxetable}

%% file: Moth.tab4.tex.rev
\begin{deluxetable}{cccccccc}
\tabletypesize{\small}
\tablecaption{2.0 $\leq$ z $\leq$ 3.5 Galaxies}
\tablewidth{0pt}
\tablehead{Id\tablenotemark{a} & z\tablenotemark{b}  &
$M_H$\tablenotemark{c} & n($H_{160}$)\tablenotemark{d} & Class\tablenotemark{e} &
R($H_{160}$) & R($H_{160}$) &
$\Delta(U_{218}-U_{300})_o$/$\Delta$log(r)\tablenotemark{f}  \\
 & &  & & & (arcsec) & (kpc) &   
}

\startdata
2-82 & 2.267 & 23.81  & 0.645 $\pm$ 0.002 & Exp & 0.396 & 2.573 & 0.4632 $\pm$ 0.0476 \\
2-239 & 2.427 & 24.58 & 0.974 $\pm$ 0.004 & - & 0.368 & 2.361 & 0.2888 $\pm$
0.0451 \\
2-454 & 2.009 & 23.55 & 0.629 $\pm$ 0.001 & Exp & 0.376 & 2.488 & 0.2039 $\pm$ 0.0382 \\
2-449 & 2.005 & 22.56 & 0.874 $\pm$ 0.002 & deV & 0.336 & 2.223 & 0.4858 $\pm$
0.0388 \\
2-585.1 & 1.980 & 24.39 & 0.794 $\pm$ 0.004 & deV & 0.288 & 1.905 & -0.0803 $\pm$
0.0536 \\
2-585.11 & 1.980 & 22.52 & 0.766 $\pm$ 0.002 & Exp & 0.592 & 3.917 & 0.0651
$\pm$ 0.0382 \\
2-76 & 3.160  & 25.37 & - & - & 0.304 & 1.826 & -0.1846 $\pm$ 0.0504 \\
2-525 & 2.237 & 23.63 & 0.840 $\pm$ 0.003 & deV & 0.428 & 2.787 & 0.4064 $\pm$
0.0395 \\
2-901 & 3.181 & 24.19 & 0.735 $\pm$ 0.002 & deV & 0.368 & 2.206 & 0.1620 $\pm$
0.0423 \\
2-565 & 3.162 & 24.60 & - & - & 0.308 & 1.849 & 0.3614 $\pm$ 0.0413 \\
2-903 & 2.233 & 24.04 & 0.706 $\pm$ 0.001 & deV & 0.259 & 1.694 & 0.4189 $\pm$
0.0467 \\
2-637 & 3.368 & 24.98 & 0.688 $\pm$ 0.004 & - & 0.212 & 1.249 & 0.7873 $\pm$
0.0451 \\
1-54  & 2.929 & 23.10 & -  & O & 0.388 & 2.382 & -0.1806 $\pm$ 0.0423 \\
2-643 & 2.991 & 24.41 & - & O & 0.339 & 2.075 & 0.5900 $\pm$ 0.0510 \\
2-591 & 2.489 & 24.04 & 0.787 $\pm$ 0.002 & deV & 0.244 & 1.557 & 0.1790 $\pm$
0.0450 \\
4-639 & 2.591 & 24.66 & 0.764 $\pm$ 0.003 & - & 0.248 & 1.569 & 0.5355 $\pm$
0.0393 \\
4-52 & 2.931 & 23.58 & - & O & 0.488 & 2.995 & -0.2922 $\pm$ 0.0392 \\
3-118 & 2.232 & 23.52 & 0.865 $\pm$ 0.002 & deV  & 0.248 & 1.615 & 0.3223 $\pm$
0.0439 \\
3-243 & 3.233 & 25.17 & 0.769 $\pm$ 0.006 & - & 0.204 & 1.217 & 0.4658 $\pm$
0.0480 \\
4-445 & 2.500 & 22.56 & 0.913 $\pm$ 0.003 & deV & 0.336 & 2.183 & -0.0043 $\pm$
0.0400 \\
4-289 & 2.969 & 25.63 & - & - & 0.264 & 1.615 & 0.6169 $\pm$ 0.0432 \\
4-497 & 2.800 & 25.15 & 0.709 $\pm$ 0.004 & - & 0.236 & 1.465 & -0.9074 $\pm$
0.0808 \\
4-555.12 & 2.803 & 23.12 & 0.960 $\pm$ 0.004 & deV & 0.316 & 1.962 & 0.3748 $\pm$
0.0402 \\
4-555.11 & 2.803 & 23.20 & 1.465 $\pm$ 0.006 & deV & 0.512 & 3.179 & 0.0962 $\pm$
0.0373 \\
4-363 & 2.980 & 25.06 & - & - & 0.284 & 1.735 & 1.0336 $\pm$ 0.0538 \\
2-890 & 2.305 & 25.06 & 0.758 $\pm$ 0.003 & - & 0.284 & 1.840 & 0.4384 $\pm$
0.0407 \\
2-949 & 2.579 & 25.04 & 0.752 $\pm$ 0.003 & - & 0.268 & 1.698 & 0.3643 $\pm$
0.0443 \\
2-746 & 3.074 & 26.25 & - & - & 0.328 & 1.986 & 0.4021 $\pm$ 0.0797 \\
2-313 & 3.417 & 25.37 & 0.783 $\pm$ 0.004 & - & 0.224 & 1.313 & 0.3740 $\pm$
0.0649 \\
2-122 & 3.066 & 25.78 & 0.505 $\pm$ 0.002 & - & 0.220 & 1.333 & -0.0774 $\pm$
0.0433 \\
2-131 & 3.184  & 24.89 & - & - & 0.300 & 1.798 & 0.3513 $\pm$ 0.0400 \\
3-675 & 3.531 &  25.31 & 0.451 $\pm$ 0.002 & - & 0.296 & 1.717 & -0.0132 $\pm$ 0.0423 \\
3-748 & 2.521  & 25.26 & 0.790 $\pm$ 0.006 & - & 0.240 & 1.528 & -0.2093 $\pm$
0.0573 \\
4-83  & 2.930  & 25.63 & -   & - & 0.280 & 1.719 & 0.2479 $\pm$ 0.0500 \\
4-245  & 3.155 & 24.73 & 0.767 $\pm$ 0.004 & - & 0.468 & 2.813 & -0.4285 $\pm$
0.0528 \\
3-813  & 3.149 & 25.77 & 0.827 $\pm$ 0.009 & - & 0.252 & 1.515 & 0.3719  $\pm$
0.0433 \\
3-875  & 2.330 & 23.17 & 0.510 $\pm$ 0.001 & Exp & 0.460 & 2.974 & 0.0438  $\pm$ 0.0391 \\
3-916  & 2.159 & 23.85 & 0.808 $\pm$ 0.003 & deV & 0.284 & 1.860 & -0.0544 $\pm$ 0.0482 \\
4-421 & 2.119 & 25.67 & 0.772 $\pm$ 0.007 & - & 0.232 & 1.524 & 0.5711   $\pm$ 0.0497 \\
2-59  & 2.255 & 27.37 & 0.574 $\pm$ 0.005 & - & 0.152 & 0.989 & 1.1681   $\pm$
0.1169 \\
2-755 & 1.970 & 25.17 & - & - & 0.376 & 2.493 & 0.3517	 $\pm$ 0.0400 \\
2-180 & 2.035 & 25.77 & 0.680 $\pm$ 0.004 & - & 0.312 & 2.061 & 1.0173   $\pm$ 0.0402 \\
2-693 & 2.012 & 23.79 & - & O & 0.344 & 2.275 & 0.1145	 $\pm$ 0.0408 \\
2-321 & 2.503 & 24.49 & 0.644 $\pm$ 0.002 & deV & 0.224 & 1.428 & 0.4166   $\pm$ 0.0415 \\
2-790 & 2.025 & 25.47 & 0.622 $\pm$ 0.003 & - & 0.260 & 1.718 & 0.8217   $\pm$
0.0652 \\
2-496 & 2.936 & 25.81 & 0.728 $\pm$ 0.004 & - & 0.208 & 1.276 & -0.3807  $\pm$
0.1180 \\
2-973 & 2.775 & 25.90 & - & - & 0.268 & 1.668 & -0.1910 $\pm$ 0.0485 \\
2-392 & 2.206 & 25.38 & - & - & 0.288 & 1.880 & 0.4641	 $\pm$ 0.1094 \\
2-347 & 2.307 & 25.30 & 0.525 $\pm$ 0.003& - & 0.236 & 1.529 & 0.9703   $\pm$
0.0532 \\
1-67 & 2.124 & 25.56 & 0.571 $\pm$ 0.004 & - & 0.240 & 1.576 & -0.4562   $\pm$
0.1162 \\
\enddata
\tablenotetext{a}{ Ids are from Williams et al. 1996}
\tablenotetext{b}{ Redshifts are from Steidel et al. 1996, Lowenthal et al. 1997, and
Budav\'ari et al. 2000}  
\tablenotetext{c}{ F160W AB mag within $\sim$ $26 \ mag/arcsec^{2}$}
\tablenotetext{d}{ S\'ersic index in rest-frame B}
\tablenotetext{e}{ deV= de Vaucouleurs, Exp = Exponential, Int = Intermediate, O = Other}
\tablenotetext{f}{ Rest-frame $(UV_{218}-U_{300})_o$ color gradient within $\sim$ $28 \
mag/arcsec^{2}$}

\end{deluxetable}

%% file: Moth.tab5.tex.rev
\begin{deluxetable}{cccccc}
\tabletypesize{\small}
\tablecaption{Morphological Classification}
\tablewidth{0pt}
\tablehead{Redshift & Total &  de Vaucouleurs&  Exponentials
& Intermediates & Others 
}
\startdata

0.5 $\leq$ z $\leq$ 1.2 & 32 & 28 $\pm$ 9\% (9 $\pm$ 3)  & 53 $\pm$ 13\% (17 $\pm$ 4)  & 13 $\pm$ 6\% (4
$\pm$ 2) &
6 $\pm$ 3\% (2 $\pm$ 1) \\
2.0 $\leq$ z $\leq$ 3.5 & 20 & 60 $\pm$ 15\% (12 $\pm$ 3) & 20 $\pm$ 10\% (4 $\pm$ 2) & 0 ( 0 ) & 20 $\pm$ 10\% (4 $\pm$ 2) \\

\enddata
\end{deluxetable}

%% file: Moth.figcap.tex.rev
\figcaption
{$I_{814}$ surface brightness profiles for our intermediate redshift sample.  The dotted line
represents a
generalized exponential fit to the profile.  Included in the plots are the galaxy ID from the
Williams et al. 1996 catalog and the value of the S\'ersic index (n).  
Those galaxies with no S\'ersic index indicated have irregular structure precluding a fit to the profile.}
\figcaption 
{Same as Figure 1, but for $H_{160}$ in the high redshift sample.}
\figcaption
{Same as Figure 1, but for $H_{160}$ in the high redshift sample.}
\figcaption
{Half-light radii versus S\'ersic index (n) for the model disks and spheroids after they were convolved with
the $H_{160}$ PSF with
$I_{814}$= 21, 22, 23, and 24. Triangles represent
model spheroids, asterisks represent model disks, and squares represent the intermediate redshift galaxies in our sample.  
The dashed and dotted line are a weighted least-squares fit to the model spheroids and model disks respectively.  The solid lines are the
1$\sigma$ deviations of the fits.}  
\figcaption
{Half-light radii versus S\'ersic index (n) for the model disks and spheroids after they were convolved with
the $H_{160}$ PSF
with $H_{160}$= 23 and 24. Triangles represent
model spheroids, asterisks represent model disks, and squares represent the high redshift galaxies in our sample.
The dashed and dotted line are a weighted least-squares fit to the model spheroids and model disks respectively.  The solid lines are the
1$\sigma$ deviations of the fits.}    
\figcaption
{Rest-frame $(UV_{218}-U_{300})_o$ logarithmic color profiles for the intermediate redshift sample.  
The
thickness of the line represents
the
statistical errors.  The dotted line represents a weighted least-squares fit to the profile starting 
from 0.14$\arcsec$ 
outwards.  Included are the galaxy IDs and the value of the color gradient determined from the least-squares fit with their
associated errors.  The errors include the uncertainties from the fits and
the upper limit in the uncertainty due to the
systematic background
subtraction } 
\figcaption {Same as Figure 6, but for the high
redshift sample.}
\figcaption
{Same as Figure 6, but for the high redshift sample.}
\figcaption
{Redshift versus rest-frame $(UV_{218}-U_{300})_o$ color gradient with error bars indicating the statistical
errors from the least-squares fit and the upper limit in the uncertainty due to the
systematic background subtraction for all the galaxies. For
the intermediate redshift, the color gradients were determined from the
$(B_{450}-V_{606})$ color profiles whereas for the high redshift sample, the color gradients were
determined from the
$(I_{814}-J_{110})$ color profiles. }
\figcaption
{$(V_{606}-I_{814})$ logarithmic color profiles for the intermediate redshift galaxies.
The thickness of the line represents the statistical errors.  The dotted line represents a weighted least-squares 
fit to the profile starting from 0.14$\arcsec$ outwards.  Included are the galaxy IDs and the value of the color 
gradient determined from the least-squares fit with their
associated errors.  The errors include the uncertainties from the fits and
the upper limit in the uncertainty due to the
systematic background subtraction.}